\documentclass[twocolumn,preprintnumbers,floatfix,prb,superscriptaddress]{revtex4}
\usepackage{graphicx}
\usepackage{graphicx}
\usepackage{dcolumn} 
\usepackage{bm}
\usepackage{epsfig}
\begin{document} 

\title{Metal-insulator transition in layered nickelates La$_3$Ni$_2$O$_{7-\delta}$ ($\delta$=0.0, 0.5, 1)}

\author{Victor Pardo}
 \email{vpardo@ucdavis.edu}
\affiliation{Department of Physics,
  University of California, Davis, CA 95616
}

\author{Warren E. Pickett}
 \email{wepickett@ucdavis.edu}
\affiliation{Department of Physics,
  University of California, Davis, CA 95616
}


\begin{abstract}

Three low-valence layered nickelates with general formula La$_3$Ni$_2$O$_{7-\delta}$ ($\delta$=0.0, 0.5, 1) are studied by ab initio techniques. Both the insulating and metallic limits are analyzed, together with the compound at the Mott transition ($\delta$= 0.5; Ni$^{2+}$), that shows insulating behavior, with all Ni atoms in a S=1 high-spin state. The compound in the $\delta$= 1 limit (La$_3$Ni$_2$O$_6$), with mean formal valence Ni$^{1.5+}$ and hence nominally metallic, nevertheless shows a correlated {\it molecular} insulating state, produced by the quantum confinement of the NiO$_2$ bilayers and the presence of mainly d$_{z^2}$ bands (bonding-antibonding split) around the gap. The metallic compound shows a larger bandwidth of the e$_g$ states that can sustain the experimentally observed paramagnetic metallic properties. The evolution of the in-plane antiferromagnetic coupling with the oxygen content is discussed, and also the similarities of this series of compounds with the layered superconducting cuprates.

\end{abstract}

\maketitle

\section{Introduction}
Three types of metal-insulator transitions have been observed in nickel oxides. The first one occurs in the RNiO$_3$ perovskites, by changing the R cation. LaNiO$_3$ is a paramagnetic metal with enhanced correlations, but as the size of the lanthanide cation is increased, the series of RNiO$_3$ (with R=Y and from Sm to Lu in the lanthanide series) are insulators that undergo a temperature induced metal-insulator transition,\cite{rnio3_structs} with the metal-insulator transition temperature (accompanied by a transition to a low-temperature antiferromagnetic state) grows as the R-O bond length increases.\cite{rnio_goodenough_prb} This transition is accompanied of a charge-ordering (or at least charge disproportionation) phenomenon, that has been observed in various members on the insulating side of the transition.\cite{rnio_alonso_co}

A second type of metal-insulator transition has been observed very recently in ultrathin films of LaNiO$_3$ grown as multilayers on LaAlO$_3$\cite{lanio_laalo_jak} and also as thin films.\cite{lanio_films_triscone} In the multilayered case, spectroscopic studies showing a double peak structure developing in the insulating cases suggest the appearance of charge-order in the Ni layer, similar to what happens in bulk RNiO$_3$ perovskites for certain types of rare earth other than La, as discussed above. It is likely that charge order is the origin of the insulating behavior in the case of thin films as well, but no spectroscopy has been performed yet to shed light on the cause of the insulating behavior observed. A similar phenomenon of charge disproportionation has been reported very recently\cite{sto_YO_mit} at the interface between two SrTiO$_3$ thin films when a monolayer of RO is grown at the interface, where R can be various lanthanides. It turns out that for the La case the interface is metallic, but for Y and the other lanthanides it is insulating, due to a similar phenomenon of charge ordering as in perovskite nickelates.

The third type of metal-insulator transition that has been experimentally observed in nickelates appears in the layered Ruddelsden-Popper series for n=2 and 3, i.e. La$_3$Ni$_2$O$_{8-\delta}$ and La$_4$Ni$_3$O$_{10-\delta}$.\cite{lanio_jpsj} La$_4$Ni$_3$O$_{10}$ and La$_3$Ni$_2$O$_7$ are both paramagnetic metals.\cite{lanio_jpsj,la3ni2o7_jpsj,lanio_magn_prb} However, at a certain value of $\delta$, a doping-induced metal-insulator transition takes place. Experimentally, various doping levels have been reached, up to La$_3$Ni$_2$O$_{6.35}$ and La$_4$Ni$_3$O$_{8.76}$, showing this metal-insulator transition.\cite{lanio_jpsj} In principle, a Mott-transition could take place at the Ni$^{2+}$ valence (La$_3$Ni$_2$O$_{6.5}$ or La$_4$Ni$_3$O$_9$), making those systems already insulating. This has to do with the formation of a Ni$^{2+}$:d$^8$ electronic state, which in an octahedral environment would form a S=1 state with an e$_g$$^2$ electronic configuration. The e$_g$$\uparrow$ band will be fully occupied leading to a gap opening if the Hund's coupling is larger than the bandwidth of the e$_g$ bands and no big distortion from octahedral environment occurs that would split the otherwise degenerate e$_g$ bands.  

The situation would be reminiscent of what happens on La$_2$NiO$_4$, which also, by electron count, has nominally Ni$^{2+}$ cations. The material shows insulating behavior at low temperature, and also a peculiar type of metal-insulator transition at high temperatures, that has been explained by Goodenough\cite{la2nio4_goodenough} as a splitting in the semi-itinerant d$_{x^2-y^2}$ band that opens up a gap, with the d$_{z^2}$ behaving always as localized electrons, according to his picture. How far the insulating region reaches in $\delta$ values will depend on how quick metallicity is reached when the insulating system is doped with either holes or electrons. Experiments show that even La$_3$Ni$_2$O$_{6.84}$ (formal valence Ni$^{2.34+}$) is an insulating compound,\cite{lanio_jpsj} suggesting that the insulating phase is difficult to destroy by doping, more difficult than in the case of cuprates, probably due to the relatively small polaron size in doped nickelates. Below, we will discuss the distinctions compared to layered cuprates that might help explaining this difference in behavior. The use of these low-valence layered nickelates has been recently suggested as a possible way to mimic the electronic structure of superconducting cuprates.\cite{giniyat,lanio_khaliullin}

Other Ni$^{2+}$ insulating compounds have been found, e.g. LaNiO$_{2.5}$, with a superstructure based on the perovskite unit cell.\cite{lanio25_alonso}
When lowering the valence below Ni$^{2+}$, the layered nickelates that have been synthesized remain insulating. Experiments on recently synthesized La$_4$Ni$_3$O$_8$\cite{lanio_2006,lanio_2007,lanio_curro_1,lanio_polt1,lanio_curro_2,la4ni3o8_vpardo} show it is an insulator with antiferromagnetic (AF) order that leads to a peculiarity in the susceptibility at high fields. In the case of La$_4$Ni$_3$O$_8$, an unusual type of molecular insulating state has been used to understand its electronic and magnetic properties,\cite{la4ni3o8_vpardo} where the quantum confinement of the NiO$_2$ trilayers leads to a bonding-antibonding splitting that creates an insulating state with a correlated molecular origin. Below, we will see how a similar picture applies to La$_3$Ni$_2$O$_6$, which is also insulating, but no AF order has been observed down to 4 K.\cite{la3ni2o6_prl}

Even lower Ni valence layered compounds exist and again a crossover from insulating to metallic state occurs as valence is reduced. The infinite-layer LaNiO$_2$ has been synthesized, with single-valent Ni$^+$:d$^9$ cations,\cite{lanio2_jacs} showing a paramagnetic metallic behavior. Electronic structure calculations show the origin of the metallicity of this compound and study the differences with the isoelectronic parent compounds of the high-temperature superconducting cuprates.\cite{lanio2_wep}



\begin{figure*}[ht]
\begin{center}
\includegraphics[width=0.6\columnwidth,draft=false]{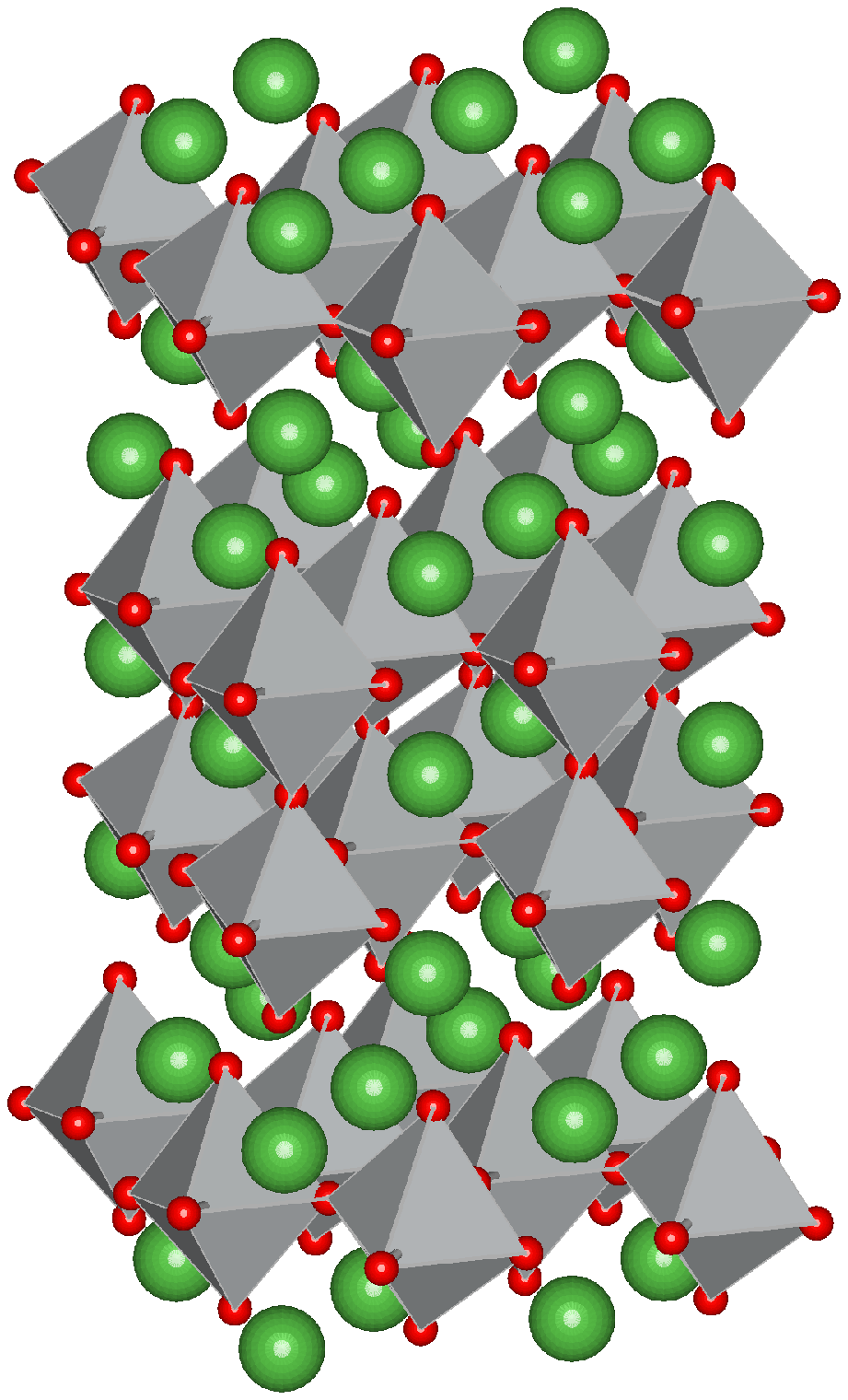}
\includegraphics[width=0.75\columnwidth,draft=false]{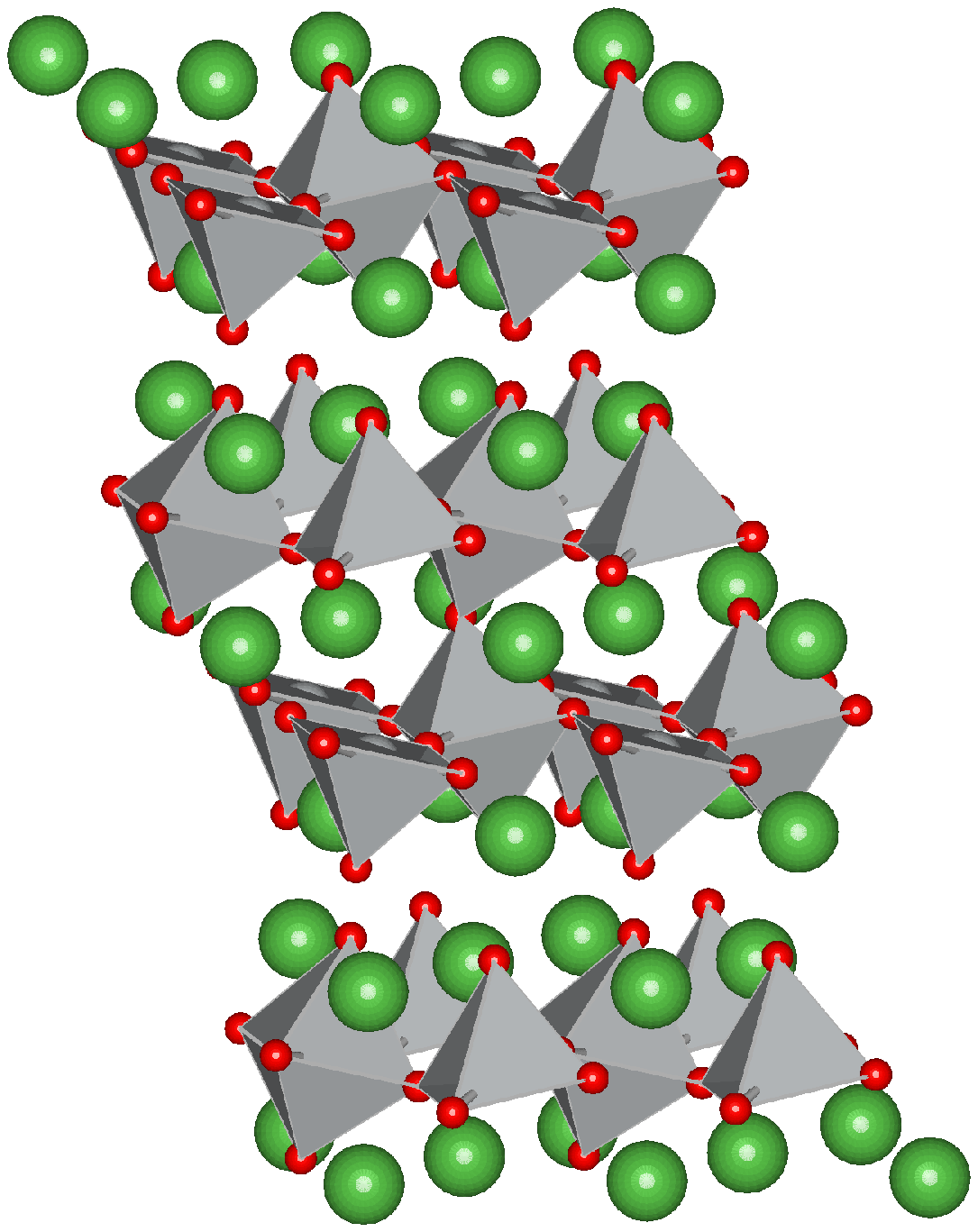}
\includegraphics[width=0.68\columnwidth,draft=false]{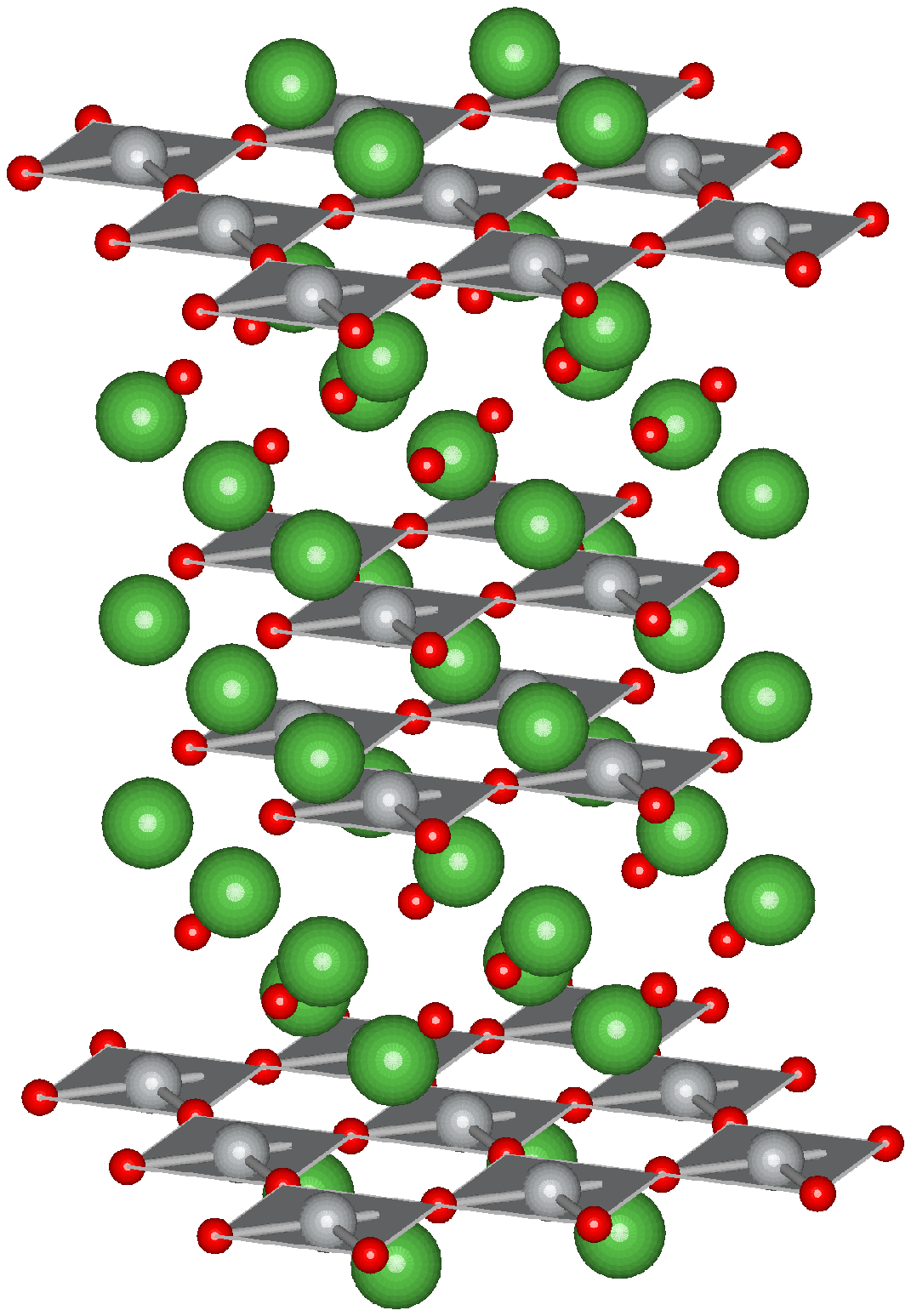}
\caption{(color online) Structure of the layered nickelates, removing oxygen from left to right: La$_3$Ni$_2$O$_7$, the metallic limit on the left side; La$_4$Ni$_3$O$_{6.5}$, the Mott-transition oxygen content in the middle; and La$_4$Ni$_3$O$_{6}$, the molecular correlated insulating limit on the right. Large (green) sphers denote La, oxygen lies at the vertices of the octahedra, pyramids, and squares.}\label{structs}
\end{center}
\end{figure*}

\section{Structure}

We have studied the n= 2 members of the Ruddelsden-Popper\cite{rp_series_2} series La$_{n+1}$Ni$_n$O$_{3n+1}$, with reduced oxygen content, to explore the regime of Ni valencies around 2+. These structures are characterized by the presence of two NiO$_2$ layers separated by a block of fluorite La/O$_2$/La layer that produces the confinement of the NiO$_2$ bilayer in the structure (see Fig. \ref{structs}) and determines the valence.

As the oxygen content is varied, so does the Ni coordination. In the insulating phase La$_3$Ni$_2$O$_6$ (right panel of Fig. \ref{structs}), the Ni environments are square planar. However, in the metallic La$_3$Ni$_2$O$_7$, every Ni is in an octahedral oxygen environment (left panel of Fig. \ref{structs}). On going from one oxygen content in the metallic limit to the other in the insulating side of the transition, apical oxygens need to be removed, that would otherwise be in the Ni bi-layer. In the intermediate case, La$_3$Ni$_2$O$_{6.5}$, we see that the reduction of oxygens leads to a mixture of octahedral and square-pyramidal environments. Structures used for the calculations in the insulating and metallic limits were taken from Refs. \onlinecite{la3ni2o6_struct,la3ni2o7_struct}. For the relaxation of the intermediate structure, we have considered the possible apical oxygen removals and compared their total energies, leading to the structure presented in the central panel of Fig. \ref{structs}. In our calculations, we have studied an ordered alternation of octahedra and square pyramids. Such an ordering need not happen in the real compound, but it is a specific way to calculate its electronic structure using a simple unit cell. One should keep in mind that the positioning of two neighboring square pyramids along the c-axis would change the d$_{z^2}$-d$_{z^2}$ $\sigma$-bonding that has crucial consequences in the electronic structure of the compound (see below). When analyzing the possible bonding-antibonding splittings produced by such a structure, it is important to keep in mind that a random distribution of square pyramids and octahedra would modify them.

\section{Computational Details}

Our electronic structure calculations were  performed within density functional
theory\cite{dft,dft_2} using the all-electron, full potential code {\sc wien2k}\cite{wien}
based on the augmented plane wave plus local orbital (APW+lo) basis set.\cite{sjo}
The generalized gradient approximation\cite{gga} (GGA) was used for the structure optimization of the $\delta$= 0.5 compound for various ways to obtain that oxygen content starting from $\delta$= 0.
To deal with  strong correlation effects we apply the LDA+U
scheme \cite{sic1,sic2} including an on-site repulsion U and Hund's coupling J 
for the Ni $3d$ states.  Results that we present are not dependent on the specific values of U and J
within a reasonable range, and we
report results with U = 4.75 eV, J = 0.68 eV, values very similar to those determined from
constrained density functional calculations\cite{lanio_curro_1} for a similar layered nickelate.

\section{La$_3$Ni$_2$O$_6$}\label{sec_326}

\begin{figure*}[ht]
\begin{center}
\includegraphics[width=2.1\columnwidth,draft=false]{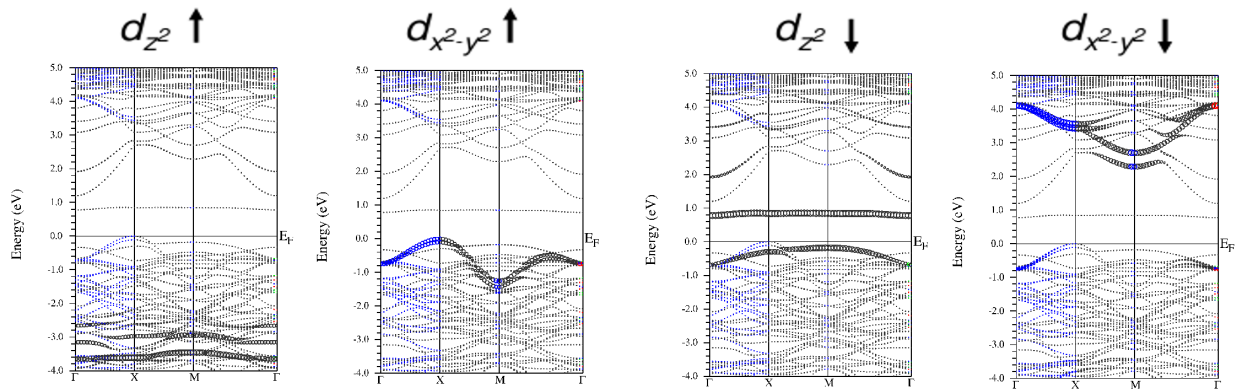}
\caption{Band structure of La$_3$Ni$_2$O$_6$, showing the fat bands for the different e$_g$ states as labelled, calculated within the LDA+U scheme, with U= 4.75 eV for Ni. The d$_{z^2}$ bands are bonding-antibonding split, while the d$_{x^2-y^2}$ bands show a larger bandwidth of about 1.5 eV.}\label{bs_326}
\end{center}
\end{figure*}

La$_3$Ni$_2$O$_6$ has been studied experimentally recently, and also using ab initio calculations.\cite{la3ni2o6_prl} The system shows largely insulating behavior that can be described by a variable range hopping model. No magnetic ordering is found down to 4 K. The calculations presented in that paper report only metallic results, hence they do not provide an understanding of the observed insulating behavior of the compound.
In our recent paper,\cite{la4ni3o8_vpardo} we describe the origin of the insulating behavior of the similar compound La$_4$Ni$_3$O$_8$ as being due to the gap opening that is produced by a bonding-antibonding splitting of the d$_{z^2}$ bands closest to the Fermi level. This comes about due to the spatial confinement of the NiO$_2$ trilayers in the compound. In the case of La$_3$Ni$_2$O$_6$, the structure presents a NiO$_2$ bilayer, but the situation becomes similar to La$_4$Ni$_3$O$_8$. Figure \ref{bs_326} shows the band structure of the compound, with the different Ni e$_g$ states highlighted, calculated within the LDA+U scheme. Due to the square planar environment of the Ni$^{1.5+}$ cations (assuming a simple electron count with the usual valences for O$^{2-}$ anions and La$^{3+}$ cations), a large crystal-field splitting inside the Ni e$_g$ doublet is expected, with the d$_{z^2}$ lying lower in energy. We can see this situation comparing the two band structures shown on the left of Fig. \ref{bs_326}. 

It is not clear what spin-state one would expect in this compound, this would depend on the difference in energy between the intra-e$_g$ crystal-field splitting and the Hund's rule coupling strength. These differences can be derived from the two central panels of Fig. \ref{bs_326}. Focusing only on the Ni atom (of which we have two in the structure) highlighted in Fig. \ref{bs_326} (the opposite spin would need to be used for the other one), if the down-spin d$_{z^2}$ band were lower in energy than the up-spin d$_{x^2-y^2}$, a low-spin state would be expected, and the d$_{x^2-y^2}$ up-spin band would be only half-filled leading to a metallic result. This low-spin state is the solution found in Ref. \onlinecite{la3ni2o6_prl}. However, our calculations yield a high-spin state to be more stable. In this case, as we see in Fig. \ref{bs_326}, the up-spin d$_{x^2-y^2}$ is comparable in energy with the down-spin d$_{z^2}$. However, there is a third energy to consider. We observe that the d$_{z^2}$ bands come in bonding-antibonding pairs, due to the spatial confinement of the NiO$_2$ bilayers and the large coupling of those orbitals along the c-axis, similar to what happens in La$_4$Ni$_3$O$_8$. That bonding-antibonding splitting leads to a gap opening around the Fermi level for this high-spin state that is favorable in energy according to our calculations. This result is consistent with the conductivity measurements performed in the compound. The fully occupied up-spin d$_{x^2-y^2}$ stabilizes an in-plane AF ordering, according to our LDA+U calculations.

A question remains as to why this system is not observed to have a long-range in-plane AF ordering. We can try to compare the bandwidths of the d$_{x^2-y^2}$ band in the case of this compound and also in La$_4$Ni$_3$O$_8$, for which we have calculations: the La$_3$Ni$_2$O$_6$ compound has a d$_{x^2-y^2}$ bandwidth of about 1.5 eV, whereas La$_4$Ni$_3$O$_8$ has a bandwidth of only 1 eV. An increase in 50\% in the bandwidth will lead to a significantly different behavior of the AF in-plane coupling. Also, in La$_2$NiO$_4$, only short-range\cite{la2nio4_pep} AF order exists, which suggests that AF in-plane ordering caused by d$_{x^2-y^2}$ orbitals is a marginal occurrence in this series of compounds. 

To quantify the evolution of antiferromagnetism, the role of the O p orbitals should be considered. The fact that in cuprates AF is stronger\cite{la4ni3o8_vpardo} and survives to larger values of doping could have to do with the O p bands playing a significant role in cuprates, but a lesser one in these layered nickelates. In the case of La$_3$Ni$_2$O$_6$, the O p bands are closer to the Fermi level (slightly more than 1 eV) than in La$_4$Ni$_3$O$_8$ (around 2 eV below the Fermi energy). Since the in-plane coupling is mediated by O p bands, differences of this kind will have implications in the different magnetic properties observed in both compounds, where electronic structure calculations predict the same magnetic ground state, with a very similar molecular correlated insulating state produced by quantum confinement.



This obvious difference (the lower position of the fully occupied O p bands compared to the cuprates) in the electronic structure between these molecular-insulating layered nickelates, both La$_3$Ni$_2$O$_6$ and La$_4$Ni$_3$O$_8$,\cite{la4ni3o8_vpardo} and the superconducting cuprates, where even Zhang-Rice singlets can occur\cite{zhang_rice} due to the position of the O p bands quite close to the Fermi level leads us to think that it could be interesting to explore the possibility of doping the oxygen sites with a larger isoelectronic anion like S. These S ions would create an anionic band with a larger bandwidth and situated higher in energy. The AF coupling in the plane could be enhanced as a result, and also the participation of the anion $p$ bands in the electronic structure of these nickelates, making them more similar to the superconducting cuprates.

Another difference lies in the fact that superconducting cuprates have a very strong Jahn-Teller coupling that produces a large elongation of the oxygen octahedra. In the case of the insulating nickelates with an electronic structure closer to that of the Cu$^{2+}$ superconducting compounds, the environment is square planar. A much smaller coupling of the electronic degrees of freedom to the lattice would be expected in the nickelates as opposed to the case of the cuprates.

One may ask: is it possible to have a pressure-induced metal-insulator transition in both La$_4$Ni$_3$O$_8$ and La$_3$Ni$_2$O$_6$? The two energies controlling the spin-state in the Ni cations are the intra-e$_g$ crystal field splitting, which can increase by reducing the $a$ and $b$ lattice parameters, and the bonding-antibonding splitting caused by the d$_{z^2}$-d$_{z^2}$ strong bonding along the $c$-axis, which increases as the $c$ parameter gets reduced. In a layered system like this, applying pressure will reduce the $c$ parameter by a larger amount than the $a,b$ in-plane lattice parameters. Pressure will move both d$_{x^2-y^2}$ bands higher in energy, and the antibonding part of the d$_{z^2}$ bands also higher in energy. It would be a question of how pressure affects the position of both bands that would determine a transition to a metallic low-spin state by applying pressure in this compound. With the experimental lattice parameters, we see from our calculations that the intra-e$_g$ crystal field splitting between d$_{z^2}$ and d$_{x^2-y^2}$ bands is about 2.8 eV, whereas the bonding-antibonding splitting between the d$_{z^2}$ up-spin bands is only 0.8 eV. Thus, if applying pressure increases the crystal-field splitting by a few tenths of an eV, a spin-state crossover may take place induced by pressure, leading to an energetically favored  metallic state.

\section{La$_3$Ni$_2$O$_7$}

\begin{figure*}[ht]
\begin{center}
\includegraphics[width=0.67\columnwidth,draft=false]{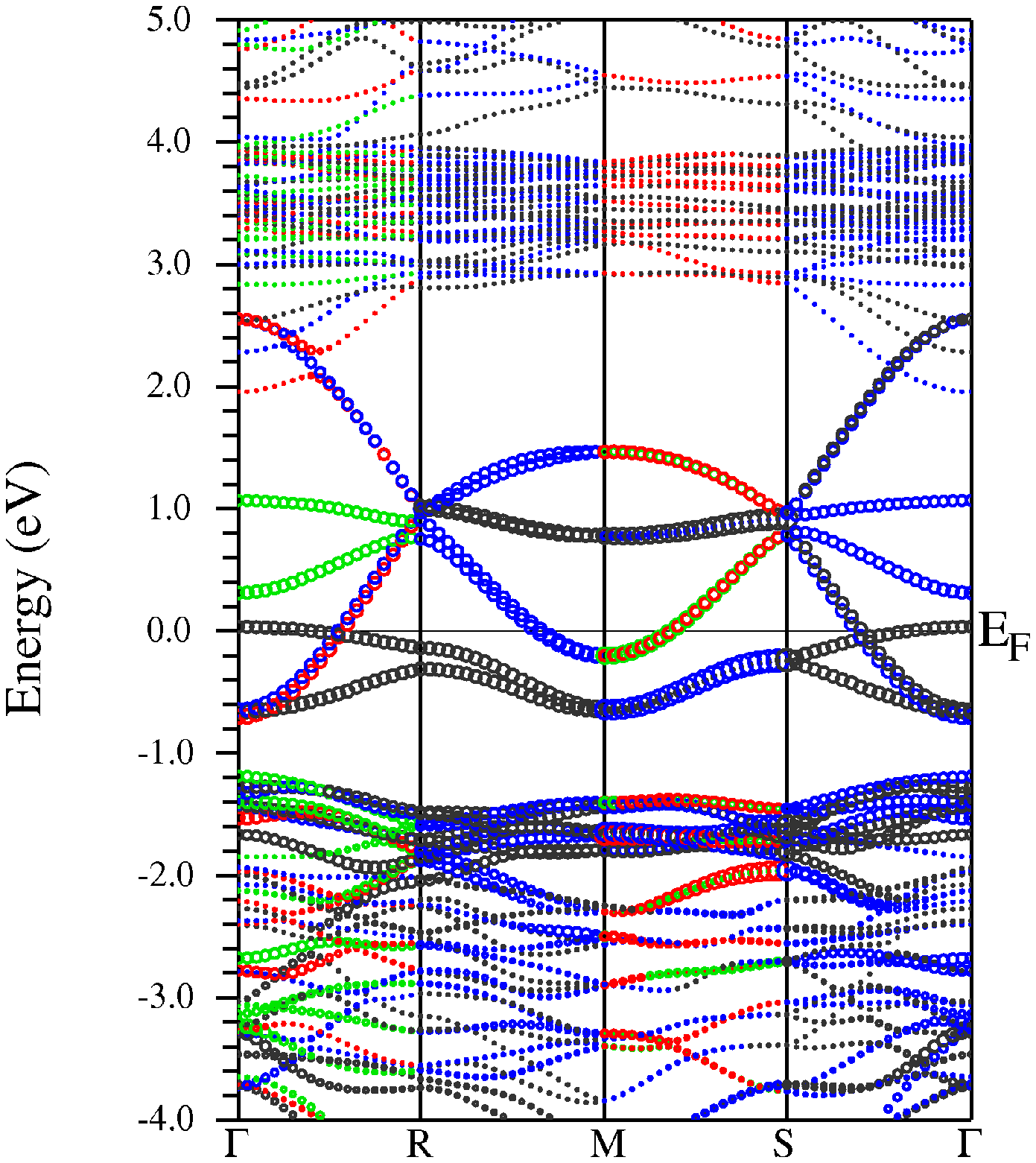}
\includegraphics[width=0.67\columnwidth,draft=false]{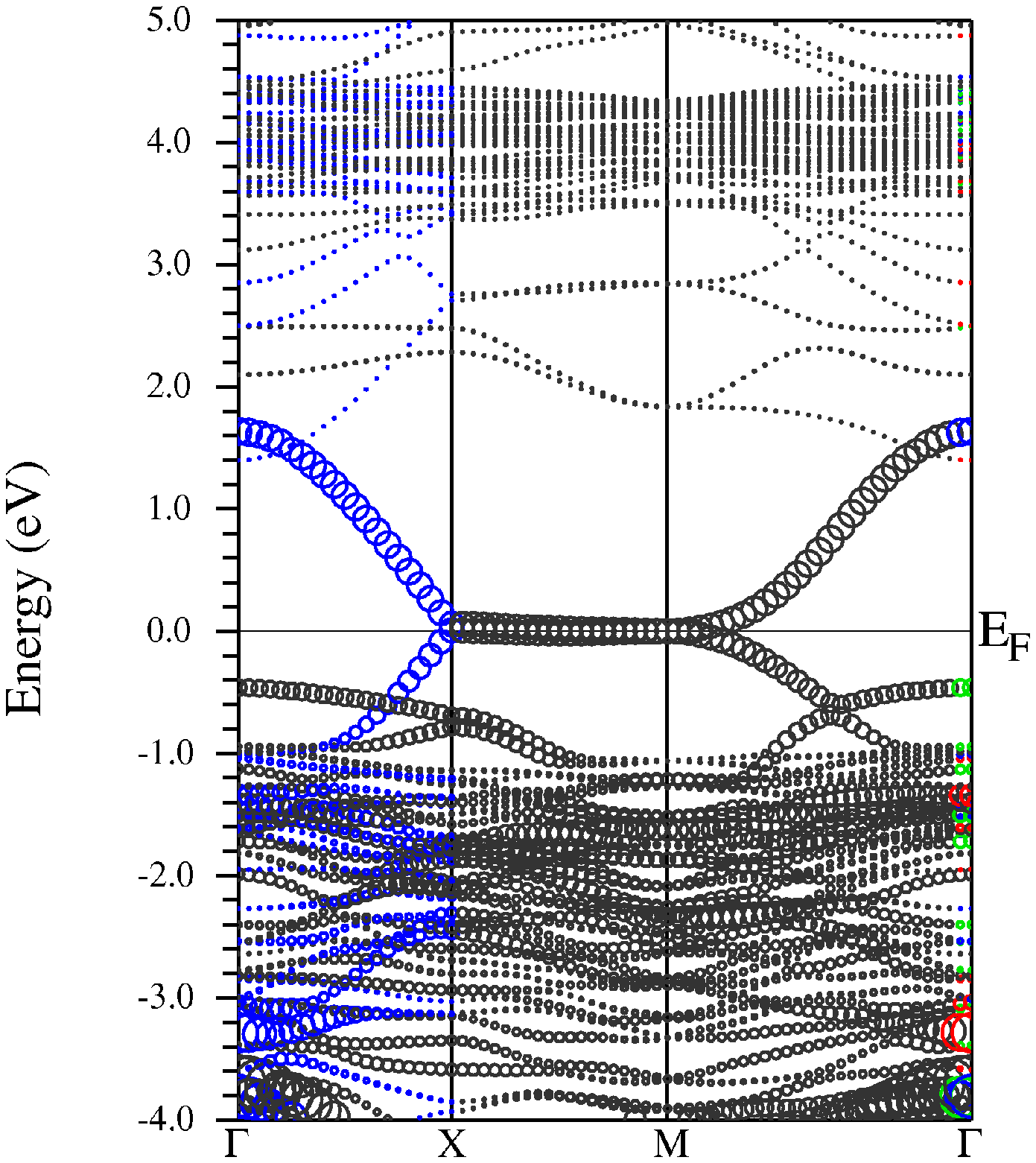}
\includegraphics[width=0.67\columnwidth,draft=false]{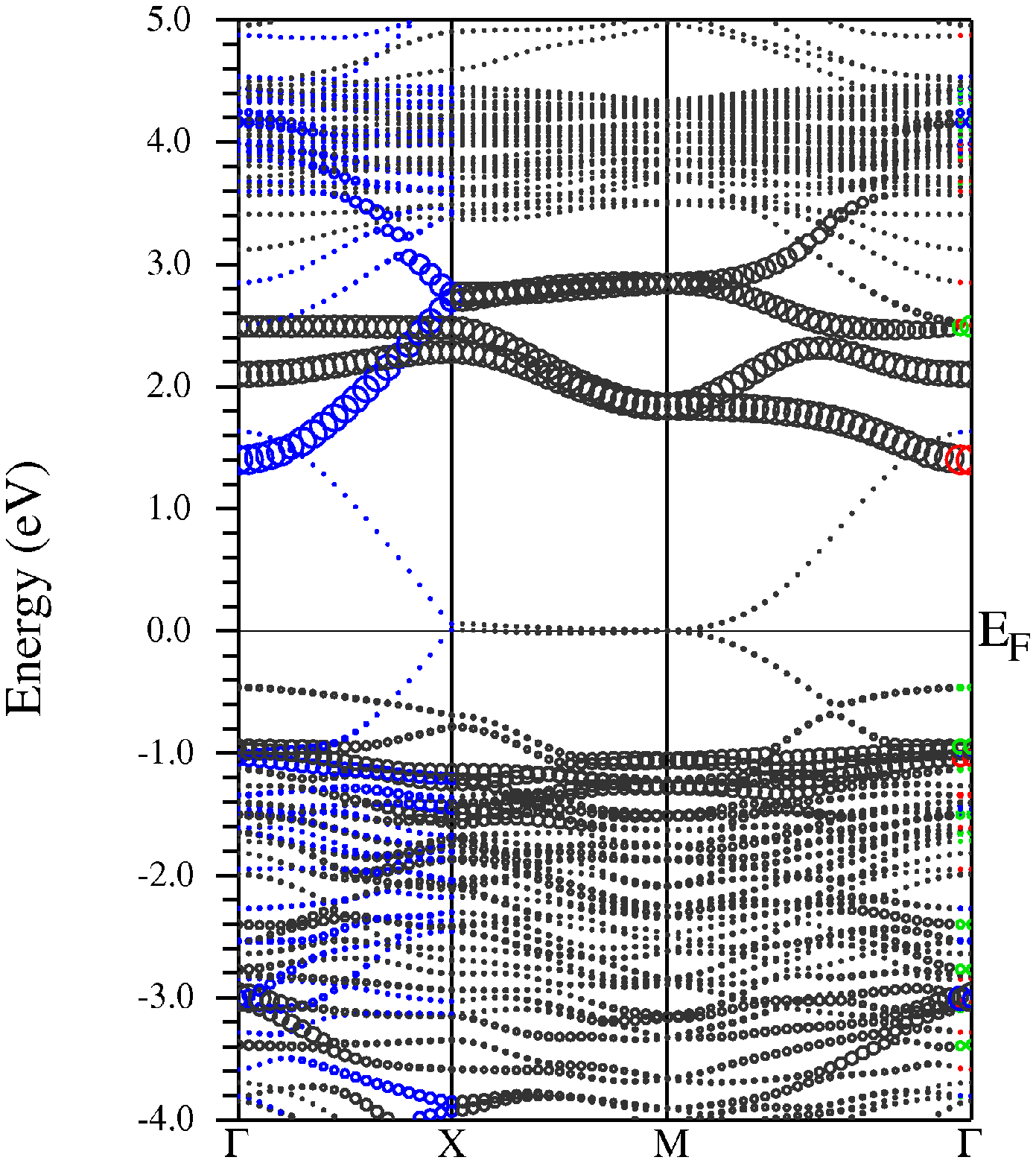}
\caption{Band structure of the La$_3$Ni$_2$O$_{7}$ in the non-magnetic phase (left panel) and the spin-up and spin-down channels of the magnetic solution (central and right panel, respectively). The calculations were performed within the LDA+U scheme, with U= 4.75 eV in Ni atoms. In all cases, the Ni states are highlighted. The e$_g$ bands occupy states close to the Fermi energy, with dispersive $d_{x^2-y^2}$ bands and relatively flat $d_{z^2}$ close in energy.}\label{bs_327}
\end{center}
\end{figure*}

The structure of La$_3$Ni$_2$O$_7$\cite{la3ni2o7_struct} is also layered, but compared to La$_3$Ni$_2$O$_6$, all the apical oxygens neighboring Ni remain, leading to all Ni atoms being in an equivalent octahedral environment. A simple electron count gives a Ni$^{2.5+}$: d$^{7.5}$ state for both Ni cations. Experiments on the compound indicate a paramagnetic metal at all temperatures. However, the formation of local moments and the appearance of dynamic AF coupling has been suggested by analyzing the susceptibility data.\cite{la3ni2o7_prb} Other studies confirm the system to be a paramagnetic metal, showing signatures of a low-dimensional localized electron system.\cite{la3ni2o7_jpsj}  The existence of a charge density wave has been suggested in the past.\cite{la3ni2o7_prb,lanio_whangbo} 

Due to its complicated magnetic and electronic structure properties, we have studied two different solutions.  One is a non-magnetic solution with an identical electronic structure in each spin channel (t$_{2g}$$^3$e$_g$$^{0.75}$), depicted in the left panel of Fig. \ref{bs_327}, with the Ni states highlighted. Around the Fermi level, the e$_g$ states can be noticed, with the  d$_{x^2-y^2}$ band also crossing the Fermi level showing a 2 eV bandwidth. The d$_{z^2}$ has a bonding and an antibonding branch due to the large interplanar coupling between those levels and the quantum confinement of the NiO$_2$ bilayers in the structure. This splitting leads to the d$_{z^2}$ bonding states to be lower in energy and more occupied than the d$_{x^2-y^2}$ bands, which would otherwise be degenerate for the two Ni sub-layers. The bonding-antibonding splitting is slightly over 1 eV, larger than in La$_3$Ni$_2$O$_6$, showing that a large hopping occurs also through an apical oxygen in the octahedral environment. The interplanar Ni-Ni distance is 3.1 \AA\ in La$_3$Ni$_2$O$_6$ and 3.9 \AA\ in La$_3$Ni$_2$O$_7$, which might seem contradictory. It is however difficult to quantify the different effects of U (the on-site Coulomb repulsion introduced in our LDA+U calculations) here due to the different occupation of that e$_g$ level compared to the case of La$_3$Ni$_2$O$_6$. In both cases, we present results with the same value of U (4.75 eV).

The other solution we have studied has a local magnetic moment for each Ni cation. This solution is lower in energy than the non-magnetic solution by a large 100 meV/Ni when LDA+U method is used with U= 4.75 eV. However, in such a metallic system the LDA+U description could be overestimating correlation effects, causing the energy difference to be too large. It is nevertheless interesting to study a magnetic solution, considering the experimental evidence for magnetic behavior.  In such a solution, the Ni$^{2.5+}$: d$^{7.5}$ cations have a spin value S=3/4, with the d$_{z^2}$ orbital fully occupied, and the d$_{x^2-y^2}$ band half-filled for each spin channel. This quarter-filling of the in-plane e$_g$ band would stabilize an in-plane AF coupling, if a localized electron picture is sufficiently valid in this itinerant compound. If correlation effects (quantified by U in our calculations) are eliminated from the calculations, the non-magnetic solution becomes more stable. The experimental evidence suggests correlated metallic behavior, which will require more sophisticated study.

\section{La$_3$Ni$_2$O$_{6.5}$}

\begin{figure*}[ht]
\begin{center}
\includegraphics[width=2.1\columnwidth,draft=false]{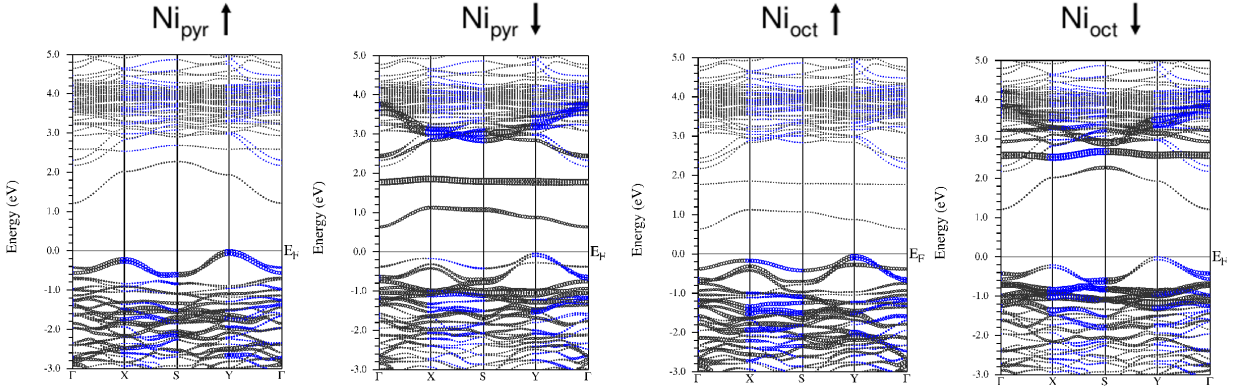}
\caption{Band structure of La$_3$Ni$_2$O$_{6.5}$, showing the fat bands for the two different Ni cations in the structure. The square pyramidal Ni is on the left, the octahedral Ni is on the right. The majority spin figures show five occupied d bands.  The minority spin figures show, above the Fermi energy, the flat $d_{z^2}$ bonding-antibonding split bands below the $d_{x^2-y^2}$ band.}\label{bs_3265}
\end{center}
\end{figure*}

The Mott metal-insulator transition in this series of compounds is expected to occur at the Ni$^{2+}$ compound. Starting to remove oxygen from the metallic La$_3$Ni$_2$O$_7$, and using the usual valences for O and La, the transition should take place at La$_3$Ni$_2$O$_{6.5}$, with Ni$^{2+}$: d$^8$ cations, if these are in a high-spin S= 1 state. The square-planar environment of the Ni cations in the insulating compound La$_3$Ni$_2$O$_{6}$ suggests that on removing oxygen from La$_3$Ni$_2$O$_7$, the first oxygens that can be removed are the apical oxygens in the NiO$_6$ octahedra.


There are four inequivalent oxygens in the structure of La$_3$Ni$_2$O$_7$, the choice of an apical oxygen that leads to a ground state is more stable by several hundreds of meV/Ni compared to removing any other oxygen. Those total energies were calculated at U= 0, with a relaxation of the atomic coordinates done within the GGA scheme. The removal of an apical oxygen leads to two different environments for the Ni cations: a NiO$_6$ distorted octahedral environment and a square pyramidal NiO$_5$ environment. Once the choice of oxygens to be removed is clarified by total energy calculations, we have further relaxed the lowest energy state within the LDA+U scheme. Additional distortions of the oxygen cages surrounding the Ni cations that occur lead to the opening of a gap at the Fermi level and the stabilization of a distinct S= 1 state in both the octahedral and the square pyramidal environments after relaxation.

We can describe the slightly different electronic structures of the two inequivalent Ni atoms in the, respectively, square pyramidal and octahedral environments with the aid of the fat bands plots shown in Fig. \ref{bs_3265}. Both cations are in an S=1 state (t$_{2g}$$^6$e$_g$$^2$). The main difference between them is in the position of the d$_{z^2}$ band. In the case of the octahedral environment, it is in the same energy window as the d$_{x^2-y^2}$ band, particularly easy to see in the down spin channel on the third panel of Fig. \ref{bs_3265}, the flat band highlighted just below the Fermi level. In the case of the square pyramidal environment, the absence of one of the apical oxygens leads the d$_{z^2}$ band lying lower in energy than the d$_{x^2-y^2}$ band, due to the higher repulsion with oxygen undergone by the latter. The bandwidth of the band that transfers the in-plane AF coupling observed to be the ground state for this compound is small compared to that of the La$_3$Ni$_2$O$_6$, being only 0.6 eV. This small bandwidth could indicate that the in-plane AF ordering is even less likely in this borderline compound than in the more insulating ones, where it is also only marginally stable. Again, the O p bands that mediate the coupling are several eV below the Fermi level, weakening the superexchange process.


\section{Summary}

In this paper we have presented a study of the evolution of the electronic structure in low-valence layered nickelates with varying oxygen content (hence, Ni valence).  The system La$_3$Ni$_2$O$_{7-\delta}$, with $\delta$= 0.0, 0.5 and 1, has been studied. A metal-insulator transition takes place between a molecular insulating state for $\delta$= 1 to a paramagnetic metal and $\delta$= 1, via a Mott-transition compound formed by Ni$^{2+}$: d$^8$ cations, with half-filled e$_g$ bands. The main difference with layered cuprates lies in the absence of the O p bands close to the Fermi level; in nickelates, only Ni d states are close to the Fermi level. One consequence of this is a less strong in-plane AF coupling between Ni $d_{x^2-y^2}$ orbitals, and also a different range of existence of the AF ordering and the insulating phase with respect to oxygen doping. The insulating properties of La$_3$Ni$_2$O$_6$ are understood as a process due to quantum confinement of the NiO$_2$ bilayer that leads
to a bonding-antibonding splitting between the Ni d$_{z^2}$ orbitals. 
The resulting gap gives the insulating behavior.

\section{Acknowledgments}
The authors have benefited from discussions with D. I. Khomskii. This project was supported by Department of Energy grant DE-FG02-04ER46111.


\begin{thebibliography}{36}
\expandafter\ifx\csname natexlab\endcsname\relax\def\natexlab#1{#1}\fi
\expandafter\ifx\csname bibnamefont\endcsname\relax
  \def\bibnamefont#1{#1}\fi
\expandafter\ifx\csname bibfnamefont\endcsname\relax
  \def\bibfnamefont#1{#1}\fi
\expandafter\ifx\csname citenamefont\endcsname\relax
  \def\citenamefont#1{#1}\fi
\expandafter\ifx\csname url\endcsname\relax
  \def\url#1{\texttt{#1}}\fi
\expandafter\ifx\csname urlprefix\endcsname\relax\def\urlprefix{URL }\fi
\providecommand{\bibinfo}[2]{#2}
\providecommand{\eprint}[2][]{\url{#2}}

\bibitem[{\citenamefont{Garcia-Munoz et~al.}(1992)\citenamefont{Garcia-Munoz,
  Rodriguez-Carvajal, Lacorre, and Torrance}}]{rnio3_structs}
\bibinfo{author}{\bibfnamefont{J.~L.} \bibnamefont{Garc\'{i}a-Mu\~noz}},
  \bibinfo{author}{\bibfnamefont{J.~L.} \bibnamefont{Rodr\'{i}guez-Carvajal}},
  \bibinfo{author}{\bibfnamefont{P.}~\bibnamefont{Lacorre}}, \bibnamefont{and}
  \bibinfo{author}{\bibfnamefont{J.~B.} \bibnamefont{Torrance}},
  \bibinfo{journal}{Phys. Rev. B} \textbf{\bibinfo{volume}{46}},
  \bibinfo{pages}{4414} (\bibinfo{year}{1992}).

\bibitem[{\citenamefont{Zhou et~al.}(2003)\citenamefont{Zhou, Goodenough, and
  Dabrowski}}]{rnio_goodenough_prb}
\bibinfo{author}{\bibfnamefont{J.~S.} \bibnamefont{Zhou}},
  \bibinfo{author}{\bibfnamefont{J.~B.} \bibnamefont{Goodenough}},
  \bibnamefont{and}
  \bibinfo{author}{\bibfnamefont{B.}~\bibnamefont{Dabrowski}},
  \bibinfo{journal}{Phys. Rev. B} \textbf{\bibinfo{volume}{67}},
  \bibinfo{pages}{020404} (\bibinfo{year}{2003}).

\bibitem[{\citenamefont{Alonso et~al.}(1999)\citenamefont{Alonso, Garcia-Munoz,
  Fernandez-Diaz, Aranda, Martinez-Lope, and Casais}}]{rnio_alonso_co}
\bibinfo{author}{\bibfnamefont{J.~A.} \bibnamefont{Alonso}},
  \bibinfo{author}{\bibfnamefont{J.~L.} \bibnamefont{Garc\'{i}a-Mu\~noz}},
  \bibinfo{author}{\bibfnamefont{M.~T.} \bibnamefont{Fern\'{a}ndez-D\'{i}az}},
  \bibinfo{author}{\bibfnamefont{M.~A.~G.} \bibnamefont{Aranda}},
  \bibinfo{author}{\bibfnamefont{M.~J.} \bibnamefont{Mart\'{i}nez-Lope}},
  \bibnamefont{and} \bibinfo{author}{\bibfnamefont{M.~T.}
  \bibnamefont{Casais}}, \bibinfo{journal}{Phys. Rev. Lett.}
  \textbf{\bibinfo{volume}{82}}, \bibinfo{pages}{3871} (\bibinfo{year}{1999}).

\bibitem[{\citenamefont{Liu et~al.}(2011)\citenamefont{Liu, Okamoto, van
  Veenendaal, Kareev, Gray, Ryan, Freeland, and Chakhalian}}]{lanio_laalo_jak}
\bibinfo{author}{\bibfnamefont{J.}~\bibnamefont{Liu}},
  \bibinfo{author}{\bibfnamefont{S.}~\bibnamefont{Okamoto}},
  \bibinfo{author}{\bibfnamefont{M.}~\bibnamefont{van Veenendaal}},
  \bibinfo{author}{\bibfnamefont{M.}~\bibnamefont{Kareev}},
  \bibinfo{author}{\bibfnamefont{B.}~\bibnamefont{Gray}},
  \bibinfo{author}{\bibfnamefont{P.}~\bibnamefont{Ryan}},
  \bibinfo{author}{\bibfnamefont{J.~W.} \bibnamefont{Freeland}},
  \bibnamefont{and}
  \bibinfo{author}{\bibfnamefont{J.}~\bibnamefont{Chakhalian}},
  \bibinfo{journal}{arxiv/1101.5581}  (\bibinfo{year}{2011}).

\bibitem[{\citenamefont{Scherwitzl et~al.}(2011)\citenamefont{Scherwitzl,
  Gariglio, Gabay, Zubko, Gibert, and Triscone}}]{lanio_films_triscone}
\bibinfo{author}{\bibfnamefont{R.}~\bibnamefont{Scherwitzl}},
  \bibinfo{author}{\bibfnamefont{S.}~\bibnamefont{Gariglio}},
  \bibinfo{author}{\bibfnamefont{M.}~\bibnamefont{Gabay}},
  \bibinfo{author}{\bibfnamefont{P.}~\bibnamefont{Zubko}},
  \bibinfo{author}{\bibfnamefont{M.}~\bibnamefont{Gibert}}, \bibnamefont{and}
  \bibinfo{author}{\bibfnamefont{J.~M.} \bibnamefont{Triscone}},
  \bibinfo{journal}{arxiv/1101.5111}  (\bibinfo{year}{2011}).

\bibitem[{\citenamefont{Jang et~al.}(2011)\citenamefont{Jang, Felker, Bark,
  Wang, Niranjan, Nelson, Zhang, Su, Folkman, Baek et~al.}}]{sto_YO_mit}
\bibinfo{author}{\bibfnamefont{H.~W.} \bibnamefont{Jang}},
  \bibinfo{author}{\bibfnamefont{D.~A.} \bibnamefont{Felker}},
  \bibinfo{author}{\bibfnamefont{C.~W.} \bibnamefont{Bark}},
  \bibinfo{author}{\bibfnamefont{Y.}~\bibnamefont{Wang}},
  \bibinfo{author}{\bibfnamefont{M.~K.} \bibnamefont{Niranjan}},
  \bibinfo{author}{\bibfnamefont{C.~T.} \bibnamefont{Nelson}},
  \bibinfo{author}{\bibfnamefont{Y.}~\bibnamefont{Zhang}},
  \bibinfo{author}{\bibfnamefont{D.}~\bibnamefont{Su}},
  \bibinfo{author}{\bibfnamefont{C.~M.} \bibnamefont{Folkman}},
  \bibinfo{author}{\bibfnamefont{S.~H.} \bibnamefont{Baek}},
  \bibnamefont{et~al.}, \bibinfo{journal}{Science}
  \textbf{\bibinfo{volume}{331}}, \bibinfo{pages}{6019} (\bibinfo{year}{2011}).

\bibitem[{\citenamefont{Kobayashi et~al.}(1996)\citenamefont{Kobayashi,
  Taniguchi, Kasai, Sato, Nishioka, and Kontani}}]{lanio_jpsj}
\bibinfo{author}{\bibfnamefont{Y.}~\bibnamefont{Kobayashi}},
  \bibinfo{author}{\bibfnamefont{S.}~\bibnamefont{Taniguchi}},
  \bibinfo{author}{\bibfnamefont{M.}~\bibnamefont{Kasai}},
  \bibinfo{author}{\bibfnamefont{M.}~\bibnamefont{Sato}},
  \bibinfo{author}{\bibfnamefont{T.}~\bibnamefont{Nishioka}}, \bibnamefont{and}
  \bibinfo{author}{\bibfnamefont{M.}~\bibnamefont{Kontani}},
  \bibinfo{journal}{J. Phys. Soc. Japan} \textbf{\bibinfo{volume}{65}},
  \bibinfo{pages}{3978} (\bibinfo{year}{1996}).

\bibitem[{\citenamefont{Taniguchi et~al.}(1995)\citenamefont{Taniguchi,
  Nishikawa, Yasui, Kobayashi, Takeda, Shamoto, and Sato}}]{la3ni2o7_jpsj}
\bibinfo{author}{\bibfnamefont{S.}~\bibnamefont{Taniguchi}},
  \bibinfo{author}{\bibfnamefont{T.}~\bibnamefont{Nishikawa}},
  \bibinfo{author}{\bibfnamefont{Y.}~\bibnamefont{Yasui}},
  \bibinfo{author}{\bibfnamefont{Y.}~\bibnamefont{Kobayashi}},
  \bibinfo{author}{\bibfnamefont{J.}~\bibnamefont{Takeda}},
  \bibinfo{author}{\bibfnamefont{S.}~\bibnamefont{Shamoto}}, \bibnamefont{and}
  \bibinfo{author}{\bibfnamefont{M.}~\bibnamefont{Sato}}, \bibinfo{journal}{J.
  Phys. Soc. Japan} \textbf{\bibinfo{volume}{64}}, \bibinfo{pages}{1644}
  (\bibinfo{year}{1995}).

\bibitem[{\citenamefont{Wu et~al.}(2001{\natexlab{a}})\citenamefont{Wu,
  Neumeier, and Hundley}}]{lanio_magn_prb}
\bibinfo{author}{\bibfnamefont{G.}~\bibnamefont{Wu}},
  \bibinfo{author}{\bibfnamefont{J.~J.} \bibnamefont{Neumeier}},
  \bibnamefont{and} \bibinfo{author}{\bibfnamefont{M.~F.}
  \bibnamefont{Hundley}}, \bibinfo{journal}{Phys. Rev. B}
  \textbf{\bibinfo{volume}{63}}, \bibinfo{pages}{245120}
  (\bibinfo{year}{2001}{\natexlab{a}}).

\bibitem[{\citenamefont{Goodenough and Ramasesha}(1982)}]{la2nio4_goodenough}
\bibinfo{author}{\bibfnamefont{J.~B.} \bibnamefont{Goodenough}}
  \bibnamefont{and}
  \bibinfo{author}{\bibfnamefont{S.}~\bibnamefont{Ramasesha}},
  \bibinfo{journal}{Mat. Res. Bull.} \textbf{\bibinfo{volume}{17}},
  \bibinfo{pages}{383} (\bibinfo{year}{1982}).

\bibitem[{\citenamefont{Chaloupka and Khaliullin}(2008)}]{giniyat}
\bibinfo{author}{\bibfnamefont{J.}~\bibnamefont{Chaloupka}} \bibnamefont{and}
  \bibinfo{author}{\bibfnamefont{G.}~\bibnamefont{Khaliullin}},
  \bibinfo{journal}{Phys. Rev. Lett.} \textbf{\bibinfo{volume}{100}},
  \bibinfo{pages}{016404} (\bibinfo{year}{2008}).

\bibitem[{\citenamefont{Hansmann et~al.}(2009)\citenamefont{Hansmann, Yang,
  Toschi, Khaliullin, Andersen, and Held}}]{lanio_khaliullin}
\bibinfo{author}{\bibfnamefont{P.}~\bibnamefont{Hansmann}},
  \bibinfo{author}{\bibfnamefont{X.}~\bibnamefont{Yang}},
  \bibinfo{author}{\bibfnamefont{A.}~\bibnamefont{Toschi}},
  \bibinfo{author}{\bibfnamefont{G.}~\bibnamefont{Khaliullin}},
  \bibinfo{author}{\bibfnamefont{O.~K.} \bibnamefont{Andersen}},
  \bibnamefont{and} \bibinfo{author}{\bibfnamefont{K.}~\bibnamefont{Held}},
  \bibinfo{journal}{Phys. Rev. Lett.} \textbf{\bibinfo{volume}{103}},
  \bibinfo{pages}{016401} (\bibinfo{year}{2009}).

\bibitem[{\citenamefont{Alonso et~al.}(1997)\citenamefont{Alonso, ,
  Martinez-Lope, Garcia-Munoz, and Fernandez-Diaz}}]{lanio25_alonso}
\bibinfo{author}{\bibfnamefont{J.~A.} \bibnamefont{Alonso}}, ,
  \bibinfo{author}{\bibfnamefont{M.~J.} \bibnamefont{Mart\'{i}nez-Lope}},
  \bibinfo{author}{\bibfnamefont{J.~L.} \bibnamefont{Garc\'{i}a-Mu\~noz}},
  \bibnamefont{and} \bibinfo{author}{\bibfnamefont{M.~T.}
  \bibnamefont{Fern\'{a}ndez-D\'{i}az}}, \bibinfo{journal}{J. Phys.: Condens. Matter}
  \textbf{\bibinfo{volume}{9}}, \bibinfo{pages}{6417} (\bibinfo{year}{1997}).

\bibitem[{\citenamefont{Poltavets
  et~al.}(2006{\natexlab{a}})\citenamefont{Poltavets, Loshkin, Egami, and
  Greenblatt}}]{lanio_2006}
\bibinfo{author}{\bibfnamefont{V.~V.} \bibnamefont{Poltavets}},
  \bibinfo{author}{\bibfnamefont{K.~A.} \bibnamefont{Loshkin}},
  \bibinfo{author}{\bibfnamefont{T.}~\bibnamefont{Egami}}, \bibnamefont{and}
  \bibinfo{author}{\bibfnamefont{M.}~\bibnamefont{Greenblatt}},
  \bibinfo{journal}{Mat. Res. Bull.} \textbf{\bibinfo{volume}{41}},
  \bibinfo{pages}{955} (\bibinfo{year}{2006}{\natexlab{a}}).

\bibitem[{\citenamefont{Poltavets et~al.}(2007)\citenamefont{Poltavets,
  Loshkin, Croft, Mandal, Egami, and Greenblatt}}]{lanio_2007}
\bibinfo{author}{\bibfnamefont{V.~V.} \bibnamefont{Poltavets}},
  \bibinfo{author}{\bibfnamefont{K.~A.} \bibnamefont{Loshkin}},
  \bibinfo{author}{\bibfnamefont{M.}~\bibnamefont{Croft}},
  \bibinfo{author}{\bibfnamefont{T.~K.} \bibnamefont{Mandal}},
  \bibinfo{author}{\bibfnamefont{T.}~\bibnamefont{Egami}}, \bibnamefont{and}
  \bibinfo{author}{\bibfnamefont{M.}~\bibnamefont{Greenblatt}},
  \bibinfo{journal}{Inorg. Chem.} \textbf{\bibinfo{volume}{46}},
  \bibinfo{pages}{10887} (\bibinfo{year}{2007}).

\bibitem[{\citenamefont{Poltavets et~al.}(2010)\citenamefont{Poltavets,
  Loshkin, Nevidomskyy, Croft, Tyson, Hadermann, Tendeloo, Egami, Kotliar,
  ApRoberts-Warren et~al.}}]{lanio_curro_1}
\bibinfo{author}{\bibfnamefont{V.~V.} \bibnamefont{Poltavets}},
  \bibinfo{author}{\bibfnamefont{K.~A.} \bibnamefont{Loshkin}},
  \bibinfo{author}{\bibfnamefont{A.~H.} \bibnamefont{Nevidomskyy}},
  \bibinfo{author}{\bibfnamefont{M.}~\bibnamefont{Croft}},
  \bibinfo{author}{\bibfnamefont{T.~A.} \bibnamefont{Tyson}},
  \bibinfo{author}{\bibfnamefont{J.}~\bibnamefont{Hadermann}},
  \bibinfo{author}{\bibfnamefont{G.~V.} \bibnamefont{Tendeloo}},
  \bibinfo{author}{\bibfnamefont{T.}~\bibnamefont{Egami}},
  \bibinfo{author}{\bibfnamefont{G.}~\bibnamefont{Kotliar}},
  \bibinfo{author}{\bibfnamefont{N.}~\bibnamefont{ApRoberts-Warren}},
  \bibnamefont{et~al.}, \bibinfo{journal}{Phys. Rev. Lett.}
  \textbf{\bibinfo{volume}{104}}, \bibinfo{pages}{206403}
  (\bibinfo{year}{2010}).

\bibitem[{\citenamefont{Poltavets
  et~al.}(2009{\natexlab{a}})\citenamefont{Poltavets, Greenblatt, Fecher, and
  Felser}}]{lanio_polt1}
\bibinfo{author}{\bibfnamefont{V.~V.} \bibnamefont{Poltavets}},
  \bibinfo{author}{\bibfnamefont{M.}~\bibnamefont{Greenblatt}},
  \bibinfo{author}{\bibfnamefont{G.~H.} \bibnamefont{Fecher}},
  \bibnamefont{and} \bibinfo{author}{\bibfnamefont{C.}~\bibnamefont{Felser}},
  \bibinfo{journal}{Phys. Rev. Lett.} \textbf{\bibinfo{volume}{102}},
  \bibinfo{pages}{046405} (\bibinfo{year}{2009}{\natexlab{a}}).

\bibitem[{\citenamefont{ApRoberts-Warren
  et~al.}(2011)\citenamefont{ApRoberts-Warren, Dioguardi, Poltavets,
  Greenblatt, Klavins, and Curro}}]{lanio_curro_2}
\bibinfo{author}{\bibfnamefont{N.}~\bibnamefont{ApRoberts-Warren}},
  \bibinfo{author}{\bibfnamefont{A.~P.}~\bibnamefont{Dioguardi}},
  \bibinfo{author}{\bibfnamefont{V.~V.} \bibnamefont{Poltavets}},
  \bibinfo{author}{\bibfnamefont{M.}~\bibnamefont{Greenblatt}},
  \bibinfo{author}{\bibfnamefont{P.}~\bibnamefont{Klavins}}, \bibnamefont{and}
  \bibinfo{author}{\bibfnamefont{N.~J.} \bibnamefont{Curro}},
  \bibinfo{journal}{Phys. Rev. B} \textbf{\bibinfo{volume}{83}},
  \bibinfo{pages}{014402} (\bibinfo{year}{2011}).

\bibitem[{\citenamefont{Pardo and Pickett}(2010)}]{la4ni3o8_vpardo}
\bibinfo{author}{\bibfnamefont{V.}~\bibnamefont{Pardo}} \bibnamefont{and}
  \bibinfo{author}{\bibfnamefont{W.~E.} \bibnamefont{Pickett}},
  \bibinfo{journal}{Phys. Rev. Lett.} \textbf{\bibinfo{volume}{105}},
  \bibinfo{pages}{266402} (\bibinfo{year}{2010}).

\bibitem[{\citenamefont{Poltavets
  et~al.}(2009{\natexlab{b}})\citenamefont{Poltavets, Greenblatt, Fecher, and
  Felser}}]{la3ni2o6_prl}
\bibinfo{author}{\bibfnamefont{V.~V.} \bibnamefont{Poltavets}},
  \bibinfo{author}{\bibfnamefont{M.}~\bibnamefont{Greenblatt}},
  \bibinfo{author}{\bibfnamefont{G.~H.} \bibnamefont{Fecher}},
  \bibnamefont{and} \bibinfo{author}{\bibfnamefont{C.}~\bibnamefont{Felser}},
  \bibinfo{journal}{Phys. Rev. Lett.} \textbf{\bibinfo{volume}{102}},
  \bibinfo{pages}{046405} (\bibinfo{year}{2009}{\natexlab{b}}).

\bibitem[{\citenamefont{Hayward et~al.}(1999)\citenamefont{Hayward, Green,
  Rosseinsky, and Sloan}}]{lanio2_jacs}
\bibinfo{author}{\bibfnamefont{M.~A.} \bibnamefont{Hayward}},
  \bibinfo{author}{\bibfnamefont{M.~A.} \bibnamefont{Green}},
  \bibinfo{author}{\bibfnamefont{M.~J.} \bibnamefont{Rosseinsky}},
  \bibnamefont{and} \bibinfo{author}{\bibfnamefont{J.}~\bibnamefont{Sloan}},
  \bibinfo{journal}{J. Am. Chem. Soc.} \textbf{\bibinfo{volume}{121}},
  \bibinfo{pages}{8843} (\bibinfo{year}{1999}).

\bibitem[{\citenamefont{Lee and Pickett}(2004)}]{lanio2_wep}
\bibinfo{author}{\bibfnamefont{K.-W.} \bibnamefont{Lee}} \bibnamefont{and}
  \bibinfo{author}{\bibfnamefont{W.~E.} \bibnamefont{Pickett}},
  \bibinfo{journal}{Phys.\ Rev. B} \textbf{\bibinfo{volume}{70}},
  \bibinfo{pages}{165109} (\bibinfo{year}{2004}).

\bibitem[{\citenamefont{Ruddlesden and Popper}(1958)}]{rp_series_2}
\bibinfo{author}{\bibfnamefont{S.~N.} \bibnamefont{Ruddlesden}}
  \bibnamefont{and} \bibinfo{author}{\bibfnamefont{P.}~\bibnamefont{Popper}},
  \bibinfo{journal}{Acta Cryst.} \textbf{\bibinfo{volume}{11}},
  \bibinfo{pages}{54} (\bibinfo{year}{1958}).

\bibitem[{\citenamefont{Poltavets
  et~al.}(2006{\natexlab{b}})\citenamefont{Poltavets, Loshin, Dikmen, Croft,
  Egami, and Greenblatt}}]{la3ni2o6_struct}
\bibinfo{author}{\bibfnamefont{V.~V.} \bibnamefont{Poltavets}},
  \bibinfo{author}{\bibfnamefont{K.~A.} \bibnamefont{Loshin}},
  \bibinfo{author}{\bibfnamefont{S.}~\bibnamefont{Dikmen}},
  \bibinfo{author}{\bibfnamefont{M.}~\bibnamefont{Croft}},
  \bibinfo{author}{\bibfnamefont{T.}~\bibnamefont{Egami}}, \bibnamefont{and}
  \bibinfo{author}{\bibfnamefont{M.}~\bibnamefont{Greenblatt}},
  \bibinfo{journal}{J. Am. Chem. Soc.} \textbf{\bibinfo{volume}{128}},
  \bibinfo{pages}{9050} (\bibinfo{year}{2006}{\natexlab{b}}).

\bibitem[{\citenamefont{Ling et~al.}(1999)\citenamefont{Ling, Argyriou, Wu, and
  Neumeier}}]{la3ni2o7_struct}
\bibinfo{author}{\bibfnamefont{C.~D.} \bibnamefont{Ling}},
  \bibinfo{author}{\bibfnamefont{D.~N.} \bibnamefont{Argyriou}},
  \bibinfo{author}{\bibfnamefont{G.}~\bibnamefont{Wu}}, \bibnamefont{and}
  \bibinfo{author}{\bibfnamefont{J.~J.} \bibnamefont{Neumeier}},
  \bibinfo{journal}{J. Solid State Chem.} \textbf{\bibinfo{volume}{152}},
  \bibinfo{pages}{517} (\bibinfo{year}{1999}).

\bibitem[{\citenamefont{Hohenberg and Kohn}(1964)}]{dft}
\bibinfo{author}{\bibfnamefont{P.}~\bibnamefont{Hohenberg}} \bibnamefont{and}
  \bibinfo{author}{\bibfnamefont{W.}~\bibnamefont{Kohn}},
  \bibinfo{journal}{Phys. Rev.} \textbf{\bibinfo{volume}{136}},
  \bibinfo{pages}{B864} (\bibinfo{year}{1964}).

\bibitem[{\citenamefont{Jones and Gunnarsson}(1989)}]{dft_2}
\bibinfo{author}{\bibfnamefont{R.~O.} \bibnamefont{Jones}} \bibnamefont{and}
  \bibinfo{author}{\bibfnamefont{O.}~\bibnamefont{Gunnarsson}},
  \bibinfo{journal}{Rev. Mod. Phys.} \textbf{\bibinfo{volume}{61}},
  \bibinfo{pages}{689} (\bibinfo{year}{1989}).

\bibitem[{\citenamefont{Schwarz and Blaha}(2003)}]{wien}
\bibinfo{author}{\bibfnamefont{K.}~\bibnamefont{Schwarz}} \bibnamefont{and}
  \bibinfo{author}{\bibfnamefont{P.}~\bibnamefont{Blaha}},
  \bibinfo{journal}{Comp. Mat. Sci.} \textbf{\bibinfo{volume}{28}},
  \bibinfo{pages}{259} (\bibinfo{year}{2003}).

\bibitem[{\citenamefont{Sj{\"o}stedt et~al.}(2000)\citenamefont{Sj{\"o}stedt,
  N{\"o}rdstrom, and Singh}}]{sjo}
\bibinfo{author}{\bibfnamefont{E.}~\bibnamefont{Sj{\"o}stedt}},
  \bibinfo{author}{\bibfnamefont{L.}~\bibnamefont{N{\"o}rdstrom}},
  \bibnamefont{and} \bibinfo{author}{\bibfnamefont{D.~J.} \bibnamefont{Singh}},
  \bibinfo{journal}{Solid State Commun.} \textbf{\bibinfo{volume}{114}},
  \bibinfo{pages}{15} (\bibinfo{year}{2000}).

\bibitem[{\citenamefont{Perdew et~al.}(1996)\citenamefont{Perdew, Burke, and
  Ernzerhof}}]{gga}
\bibinfo{author}{\bibfnamefont{J.~P.} \bibnamefont{Perdew}},
  \bibinfo{author}{\bibfnamefont{K.}~\bibnamefont{Burke}}, \bibnamefont{and}
  \bibinfo{author}{\bibfnamefont{M.}~\bibnamefont{Ernzerhof}},
  \bibinfo{journal}{Phys.\ Rev. Lett.} \textbf{\bibinfo{volume}{77}},
  \bibinfo{pages}{3865} (\bibinfo{year}{1996}).

\bibitem[{\citenamefont{Anisimov et~al.}(1991)\citenamefont{Anisimov, Zaanen,
  and Andersen}}]{sic1}
\bibinfo{author}{\bibfnamefont{V.~I.} \bibnamefont{Anisimov}},
  \bibinfo{author}{\bibfnamefont{J.}~\bibnamefont{Zaanen}}, \bibnamefont{and}
  \bibinfo{author}{\bibfnamefont{O.~K.} \bibnamefont{Andersen}},
  \bibinfo{journal}{Phys. Rev. B} \textbf{\bibinfo{volume}{44}},
  \bibinfo{pages}{943} (\bibinfo{year}{1991}).

\bibitem[{\citenamefont{Ylvisaker et~al.}(2009)\citenamefont{Ylvisaker,
  Pickett, and Koepernik}}]{sic2}
\bibinfo{author}{\bibfnamefont{E.~R.} \bibnamefont{Ylvisaker}},
  \bibinfo{author}{\bibfnamefont{W.~E.} \bibnamefont{Pickett}},
  \bibnamefont{and}
  \bibinfo{author}{\bibfnamefont{K.}~\bibnamefont{Koepernik}},
  \bibinfo{journal}{Phys. Rev. B} \textbf{\bibinfo{volume}{79}},
  \bibinfo{pages}{035103} (\bibinfo{year}{2009}).

\bibitem[{\citenamefont{Fontcuberta et~al.}(1984)\citenamefont{Fontcuberta,
  Longworth, and Goodenough}}]{la2nio4_pep}
\bibinfo{author}{\bibfnamefont{J.}~\bibnamefont{Fontcuberta}},
  \bibinfo{author}{\bibfnamefont{G.}~\bibnamefont{Longworth}},
  \bibnamefont{and} \bibinfo{author}{\bibfnamefont{J.~B.}
  \bibnamefont{Goodenough}}, \bibinfo{journal}{Phys. Rev. B}
  \textbf{\bibinfo{volume}{30}}, \bibinfo{pages}{6320} (\bibinfo{year}{1984}).

\bibitem[{\citenamefont{Zhang and Rice}(1988)}]{zhang_rice}
\bibinfo{author}{\bibfnamefont{F.~C.} \bibnamefont{Zhang}} \bibnamefont{and}
  \bibinfo{author}{\bibfnamefont{T.~M.} \bibnamefont{Rice}},
  \bibinfo{journal}{Phys. Rev. B} \textbf{\bibinfo{volume}{37}},
  \bibinfo{pages}{3759} (\bibinfo{year}{1988}).

\bibitem[{\citenamefont{Wu et~al.}(2001{\natexlab{b}})\citenamefont{Wu,
  Neumeier, and Hundley}}]{la3ni2o7_prb}
\bibinfo{author}{\bibfnamefont{G.}~\bibnamefont{Wu}},
  \bibinfo{author}{\bibfnamefont{J.~J.} \bibnamefont{Neumeier}},
  \bibnamefont{and} \bibinfo{author}{\bibfnamefont{M.~F.}
  \bibnamefont{Hundley}}, \bibinfo{journal}{Phys. Rev. B}
  \textbf{\bibinfo{volume}{63}}, \bibinfo{pages}{245120}
  (\bibinfo{year}{2001}{\natexlab{b}}).

\bibitem[{\citenamefont{Seo et~al.}(1996)\citenamefont{Seo, Liang, Whangbo,
  Zhang, and Greenblatt}}]{lanio_whangbo}
\bibinfo{author}{\bibfnamefont{D.~K.} \bibnamefont{Seo}},
  \bibinfo{author}{\bibfnamefont{W.}~\bibnamefont{Liang}},
  \bibinfo{author}{\bibfnamefont{M.~H.} \bibnamefont{Whangbo}},
  \bibinfo{author}{\bibfnamefont{Z.}~\bibnamefont{Zhang}}, \bibnamefont{and}
  \bibinfo{author}{\bibfnamefont{M.}~\bibnamefont{Greenblatt}},
  \bibinfo{journal}{Inorg. Chem.} \textbf{\bibinfo{volume}{35}},
  \bibinfo{pages}{6396} (\bibinfo{year}{1996}).

\end{thebibliography}

\end{document}